# Japan's Possible Contributions for Coronagraph of the Habitable Worlds Observatory (HWO)

**Keigo Enya,[1,2] Kenta Yoneta,[1] Naoshi Murakami,[3,4,2] Jun Nishikawa,[3,4,2] Satoshi Itoh,[5] Taro Matsuo,[6] Reiki Kojima,[5] Takayuki Kotani,[3,4,2] Olivier Guyon,[7,8] Takahiro Sumi,[6] Satoshi Miyazaki,[7] Toru Yamada,[1,2] Aoi Takahashi,[1,2] Hajime Kawahara,[1] Shota Miyazaki,[1] Iona Kondo,[1] Nana Higashio,[1] Noriko Yamasaki,[1] Masayuki Kuzuhara,[3] Motohide Tamura,[3] Masahiro Ikoma,[3,4] Norio Narita,[9] Julien Lozi[78]**

[1]*Institute of Space and Astronautical Science (ISAS), Japan Aerospace Exploration Agency (JAXA), Sagamihara, Kanagawa, Japan*
[2]*The Graduate University for Advanced Studies, SOKENDAI, Hayama, Kanagawa, Japan*
[3]*Astrobiology Center (ABC), National Institutes of Natural Sciences (NINS), Tokyo, Japan*
[4]*National Astronomical Observatory of Japan (NAOJ), NINS, Tokyo, Japan*
[5]*Nagoya University, Nagoya, Japan*
[6]*Osaka University, Osaka, Japan*
[7]*Subaru Telescope, NAOJ, NINS, Hawaii, USA*
[8]*The University of Arizona, Arizona, Tucson, USA*
[9]*University of Tokyo, Tokyo, Japan*

**Abstract**

In this paper, we describe Japan's possible contributions for coronagraph of the Habitable Worlds Observatory (HWO) based on our independent study. We are considering to contribute to the HOW coronagraph by science and hardware, based on Japan's experience for the SPICA coronagraph instrument, contributions to the Nancy Grace Roman Space Telescope, and SCExAO for the Subaru telescope. Currently, hardware contributions of various scales, from large-scale to small components, are considered. As an example of the large-scale hardware case, the optical and mechanical layout of the entire infrared coronagraph is presented. Several individual high-contrast technologies are also briefly introduced, for which research is ongoing in Japan. In discussions, it is pointed out that both the inner working angle (IWA) and sensitivity are particularly critical for the NIR coronagraph. In this situation, dedicated observations of a small number of targets close to the solar system can be one of key science program in this situation, and designing consolidating science objectives, requirements, observation targets, and survey plans is important. It is essential to push the development of advanced coronagraphs that provide small IWAs. On the other hand, it is also necessary to prepare solutions that adopt more robust coronagraphs in parallel. How to coexist visible and NIR coronagraphs within constraints of volume, mass, budget etc. is an important issue. The international sharing for the coronagraph development should be carefully decided by international agreement. Although all of our studies may not be realized in contributions to the first generation of HWO instruments, we are considering Japan's multigenerational participation in the HWO to maximize outcomes of the HWO.

## 1. INTRODUCTION

The Habitable Worlds Observatory (HWO) is a proposed flagship space telescope following the James Webb Space Telescope and the Nancy Grace Roman Space Telescope (Feinberg et al. 2024). The HWO is designed to have a 6–8 m class aperture and conduct observations in the ultraviolet to near-infrared (NIR) wavelength range. The launch of the HWO is scheduled for the 2040s.

One of the HWO's primary scientific objectives is direct observation and characterization of Earth-like exoplanets (including the search for signs of life). For this purpose, the HWO will be equipped with a coronagraph instrument.

The HWO is a NASA-led project with international partnership. At this moment, contributions to the coronagraph and ultraviolet instruments of the HWO are under consideration as Japan's contribution to the HWO's hardware.

In this paper, we present Japan's possible contributions considered at this moment with various scales to the HWO coronagraph, together with Japanese background for space-borne coronagraph and discussion especially on NIR coronagraph.

## 2. JAPANESE BACKGROUND FOR SPACE-BORNE CORONAGRAPH

In this section, we present the Japanese background for spaceborne coronagraphs at the instrument or project level.

### 2.1. SPICA Coronagraph Instrument

Space Infrared Telescope for Cosmology and Astrophysics (SPICA) was a space-borne infrared telescope observatory that was first proposed in Japan, then became an ESA-JAXA mission, and finally was cancelled (Fig. 1). For SPICA, development of the SPICA coronagraph instrument (SCI) and associated technology was advanced (Fig. 1) (Enya et al. 2011). The primary objective of the SCI was direct imaging and characterization with spectroscopy of Jovian exoplanet in mid-infrared wavelength range with the contrast of $10^{-7}$. Although the specifications of the SCI are not the same as those of the HWO coronagraph, we consider that the experience and knowledge gained from developing an entire coronagraph instrument will be worthwhile for contributing to the HWO.

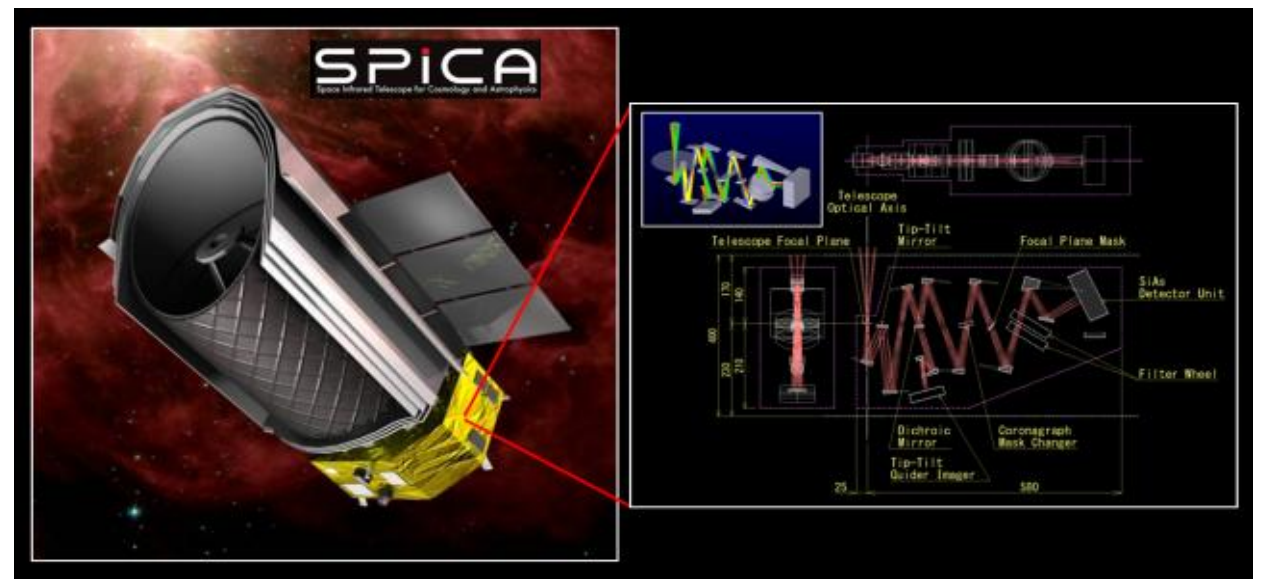

Fig. 1. — An artistic view of SPICA (left) and the design of the SPICA Coronagraph Instrument (SCI) (right).

### 2.2. Contributions to the Nancy Grace Roman Space Telescope

JAXA is contributing to the Coronagraph Instrument (CGI) of the Nancy Grace Roman Space Telescope by supplying substrate of coronagraph masks (for the hybrid Lyot coronagraph and shaped pupil coronagraph), Wollaston prisms and imaging lenses for polarimetric mode (Fig. 2) (Riggs et al. 2025; Groff et al. 2025).

In addition, approximately 20 researchers are participating in the Coronagraph Community Participation Program (CPP) through JAXA/ISAS.

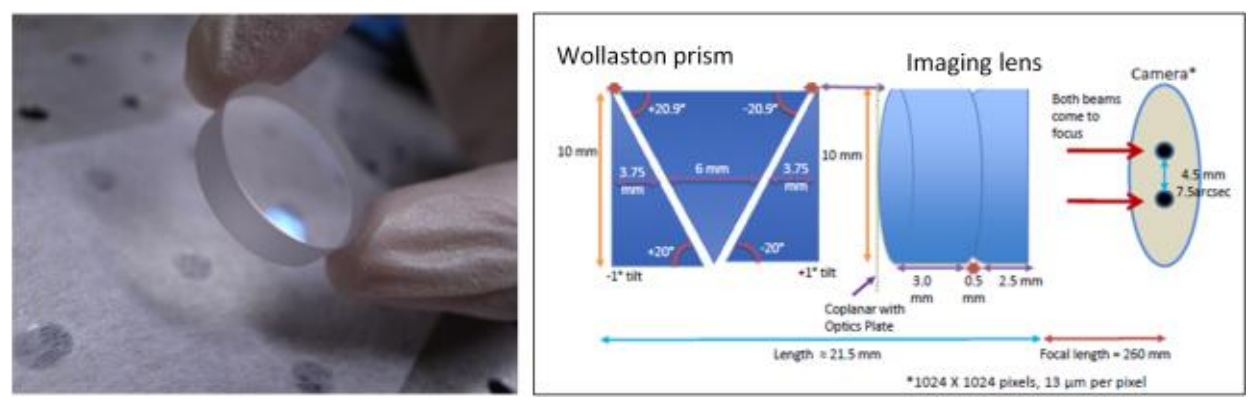

Fig. 2. — A substrate of coronagraph mask (left) and Wollaston prisms and imaging lenses for polarimetric mode (right).

### 2.3. SCExAO

Subaru Coronagraphic Extreme Adaptive Optics (SCExAO) is an astronomical instrument and development platform for high-contrast imaging on the Subaru Telescope (Fig. 3) (Jovanovic et al. 2015). The SCExAO will be used for system-level technology validation of the HWO-J coronagraph at intermediate (~$10^{-8}$) contrast, for scientific observations, and workforce development. Development of the NIR photonic (fiber-based) spectrograph concept.

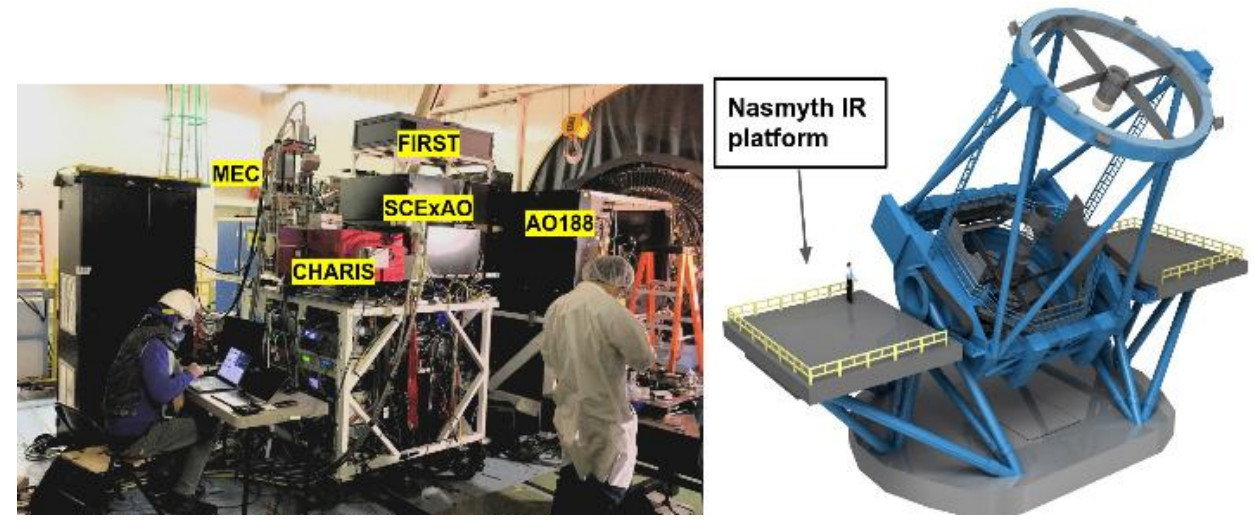

Fig. 3. — SCExAO (left) and Subaru telescope with the Nasmyth IR platform that is a place the SCExAO is to be mounted (left).

## 3. POSSIBLE CONTRIBUTIONS

### 3.1. Science contributions

Scientific contributions are essential. We consider that the ongoing contributions to the CPP for the Nancy Grace Roman Space Telescope will be a meaningful experience.

### 3.1. A large part of the hardware development

The scale of Japan's contribution should finally be determined through a well-considered international agreement. Currently, the Japan’s coronagraph team has a wish to develop a large part of the hardware. A typical large part of the hardware in a coronagraph instrument is the entire NIR channel. We performed a study for the design of the entire infrared channel in our own initiative. Ane example of optical layout of the entire NIR channel is presented in Fig. 4.

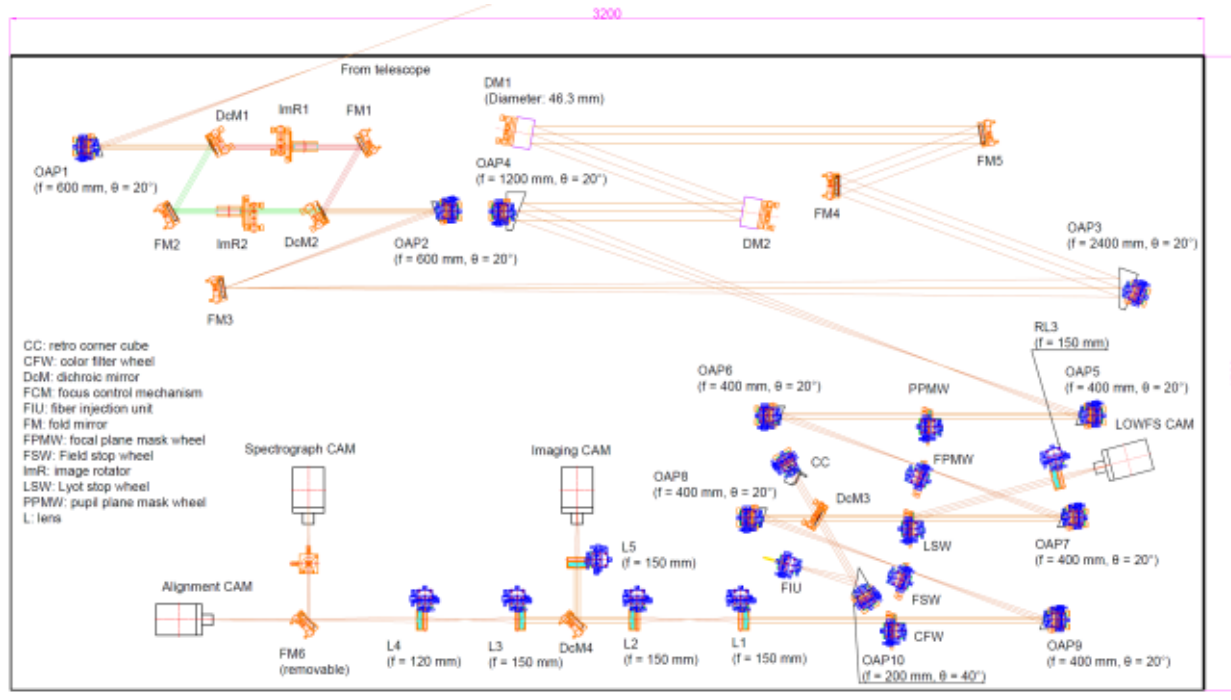

Fig. 4. — Ane example of optical layout of the entire NIR channel obtained in in a voluntarily study.

### 3.2. Development of individual high-contrast technologies

Several high-contrast technologies are currently under development in Japan. Listed below are presentations given at HWO25 (formatted as first author (affiliation): presentation title in HWO25, references). For details of these issues, please refer to their proceedings. These technologies have potential to be useful regardless of whether the scale of Japan's contributions are large or small.

① K. Yoneta (JAXA/ISAS): Development of the Speckle Area Nulling and Coherent Differential Imaging on Speckle Area Nulling (Yoneta et al. 2022).

② K. Yoneta (JAXA/ISAS): Development of High-Contrast Imaging Techniques for Exoplanets in Binary-Star Systems (Yoneta et al. 2024).

③ N. Murakami (ABC/NAOJ): Development of broadband coronagraphic phase masks and wavefront control technique (Murakami et al. (2025), in preparation [HWO25 proceedings]).

④ J. Nishikawa (NAOJ): Two-stage coronagraph for high contrast imaging (Nishikawa et al. (2025), in preparation [HWO25 proceedings]).

⑤ S. Itoh (Nagoya Univ.): Starlight Suppression: Contrast Technology & Methods (Itoh et al. 2025).

⑥ R. Kojima (Nagoya Univ.): Ultrastable Telescope and Observatory Technology, Starlight Suppression: High-contrast Imaging Technology (Kojima et al. 2025, in preparation [HWO25 proceedings]).

## 4. DISCUSSIONS

### 4.1. Inner Working Angle (ISA) and sensitivity

The IWA is an important parameter of a coronagraph. The IWA of a certain type of coronagraph is usually expressed as $\alpha\lambda/D$, where $\lambda$ is wavelength, $D$ is the optical aperture, and $\alpha$ is a constant that varies depending on the coronagraph type. Therefore, for example, the IWA of a certain type of coronagraph at a wavelength of 1.8 μm in the NIR range is three times larger (i.e., worse) than the IWA at a wavelength of 0.6 μm in the visible range.

IWA constrains the number of observable targets strongly. Table 1 shows examples of the results of evaluating the number of Earth equivalents observable in the NIR range at different IWAs using NASA Exoplanet Archive. In the case of IWA = 1, the number of targets is only nine. Moreover, if we add other conditions, e.g., excluding double stars, the number will be further reduced. However, as IWA becomes smaller, the number of targets increases rapidly. Thus, it is highly important to develop advanced coronagraphs with small IWAs.

Mennesson et al. (2024) presented the sensitivity (required exposure time) calculated under various conditions for the observation of Earth equivalents in their paper. We also performed calculations using a similar method to theirs and confirmed that there are severe sensitivity constraints and long exposure times are required for observations of Earth equivalents, especially in NIR. Therefore, it is also important to develop advanced coronagraphs with high throughput and sensitivity.

Table 1. Number of observable targets in NIR.

| IWA ($\lambda/D$) | Number of observable targets |
|---|---|
| 3 | 9 |
| 2 | 32 |
| 1 | Many |

$D$ = 6 m and $\lambda$ = 1.8 μm are assumed. The number of Earth equivalents is estimated using NASA Exoplanet Archive with requiring IWA as the only constraint (https://exoplanetarchive.ipac.caltech.edu/index.html).

### 4.2. Advanced or robust coronagraphs?

Considering the abovementioned, it is essential to conduct the development of advanced coronagraphs. On the other hand, such development is highly challenging and there is a risk that it cannot be completed. Regardless of the type of coronagraph, we consider that realizing a coronagraph with a space telescope capable of providing $10^{-10}$ contrast itself is challenging in the first place. Hence, we should also develop a relatively robust coronagraph as a backup (or baseline) plan in parallel.

### 4.3. Dedicated observations of a small number of targets close to the solar system

Under severe limits on both IWA and sensitivity, we consider that highly dedicated observations of a small number of targets very close to the solar system can be one of important science program. Multi-band imaging, spectroscopy, polarimetry, and monitoring using these methods are considered for this purpose. The exoplanet that will be the target of the observation is expected to be discovered in advance through direct imaging. The more targets there are, the better. However, even in the extreme case of just one, detailed characterization of an Earth equivalent can be valuable. This corresponds to one exoplanet system, TRAPIST-1, has played an important role in the science of transiting exoplanets. An observation program focusing on nearby targets is particularly effective in NIR but can also be productive in the visible range. This strategy is useful when the contrast achieved in coronagraph development falls slightly short of requirements. Our Japanese team is conducting studies on this topic.

### 4.4. Design consolidating science objectives, requirements, observation targets, and survey plans

As mentioned above, the development of coronagraphs targeting Earth equivalents faces strict constraints on IWA, sensitivity, and contrast. Therefore, in the design of a coronagraph, it is particularly important to handle the science objectives, requirements, observation targets, and survey plans in a consolidated manner. These should be iterated and optimized. During the design phase, observation targets and survey plans do not need to be final, but rather hypothetical. Feedback from the experiment on the performance of the coronagraph would also be valuable. This situation is common in coronagraphs regardless of the type of science program, whether it is in NIR or visible range.

### 4.5. How to coexist visible and NIR coronagraphs

There are several potentially possible solutions for where and how the visible and NIR coronagraph's functional paths should split (Fig. 5). Simplified examples for typical cases are shown. The branching for each polarization is important issue, but omitted in this figure. Among the cases shown in Fig. 5, the maximally separated case provides the best observation performance. On the other hand, advantage of the minimally separated case is simple, compact, light-weighted, and low cost.

Another similar issue is the implementation of visible and infrared coronagraphs, i.e., whether to build them on separate optical benches or use a common one.

Each of the various possible designs for these issues has its own advantages and disadvantages. Even the entire hardware for one channel is large as shown in Fig. 4. As the result, trade-off studies are necessary to finalize the design. The international sharing of hardware development will be significantly affected by the final configuration of the total design.

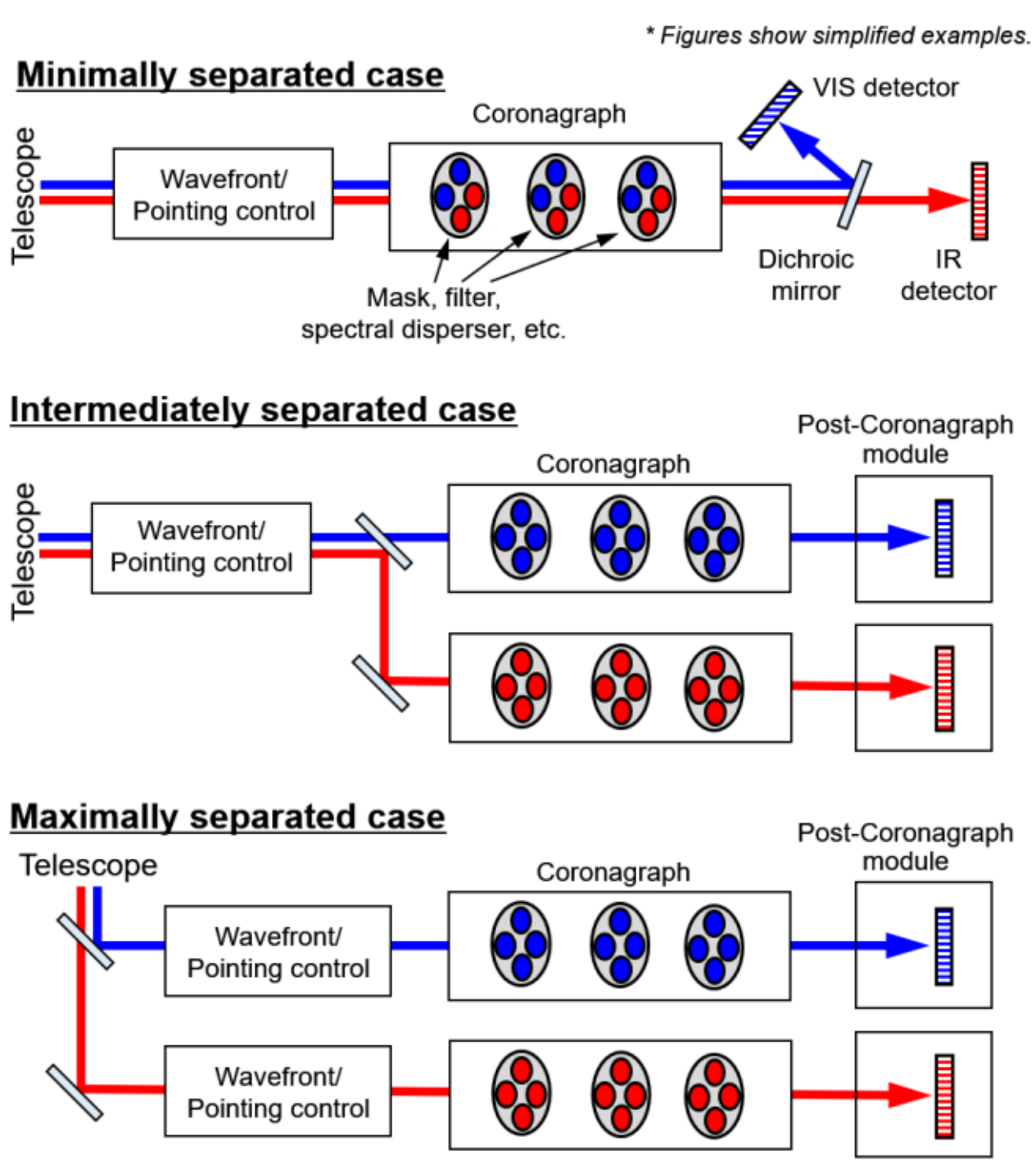

Fig. 5. — Schematic diagrams of the branch of visible and NIR optical paths. Simplified examples for typical cases are shown. The branching for each polarization is omitted.

### 4.6. Future works

Not all of the studies that the Japan team conducts from now on will directly correspond to the Japanese-responsible part of the HWO's initial instrument development. Nevertheless, such efforts could contribute to Japan playing a more prominent role in development of the HWO's instruments of the second generation and beyond, e.g., Japanese-led core science instrument of future. In order to maximize the value of the overall products of the HWO, we are considering Japan's multigenerational participation in the HWO.

## 5. FINAL REMARKS

First of all, it should be noted that the international sharing for the coronagraph development has not been decided and this paper describes our independent study. Along with optimizing and finalizing the overall architecture of the HWO coronagraph, Japan's contribution and its scale should be carefully decided by international agreement.

**Acknowledgments.**

Our participation in HWO25 is supported by the budget of the ISAS Task force established for participation in the HWO.

# Appendix

While the main body of this paper remains identical to the HWO 2025 proceedings, the appendices provide omitted details or several subsequent advancements.

## Appendix A. Optical Design

Fig. A1. Example of the NIR coronagraph optical design, generated after drawing Fig. 4.In the overall optical system layout diagram (top of Fig. A1), only the imaging part is shown as the post-coronagraph, and the spectroscopic part is omitted. This optics contains two DMs. The structure of the coronagraph part in the central section of the optics consists of multiple off-axis parabolic mirrors. Coronagraph masks, filters, and other components can be replaced using mechanical exchange mechanisms placed at the pupil and focal planes. The imaging part consists of lenses. With this simple configuration, it can cover the entire NIR wavelength range (1–1.8 μm).

This design is an example, which we carried out to understand the overall size and to gain design experience. The overall size of the optical system depends on the boundary conditions and is affected by the size of the deformable mirror. Enabling coronagraphic observations both within and beyond the habitable zone via a wide dark hole can contribute to increasing scientific productivity, despite requiring a large-format deformable mirror. Spectroscopy in a non-coronagraphic mode, achieved by removing the mask via a mechanical exchanger, enhances the scientific productivity for transiting exoplanets and other targets. For this mode, it is preferable to simultaneously observe the same target over a wide wavelength range.

This design is currently a work in progress. It will become more specific once the requirements, boundary conditions, constraints, and the architecture for the coexistence of the VIS and NIR coronagraphs are determined.

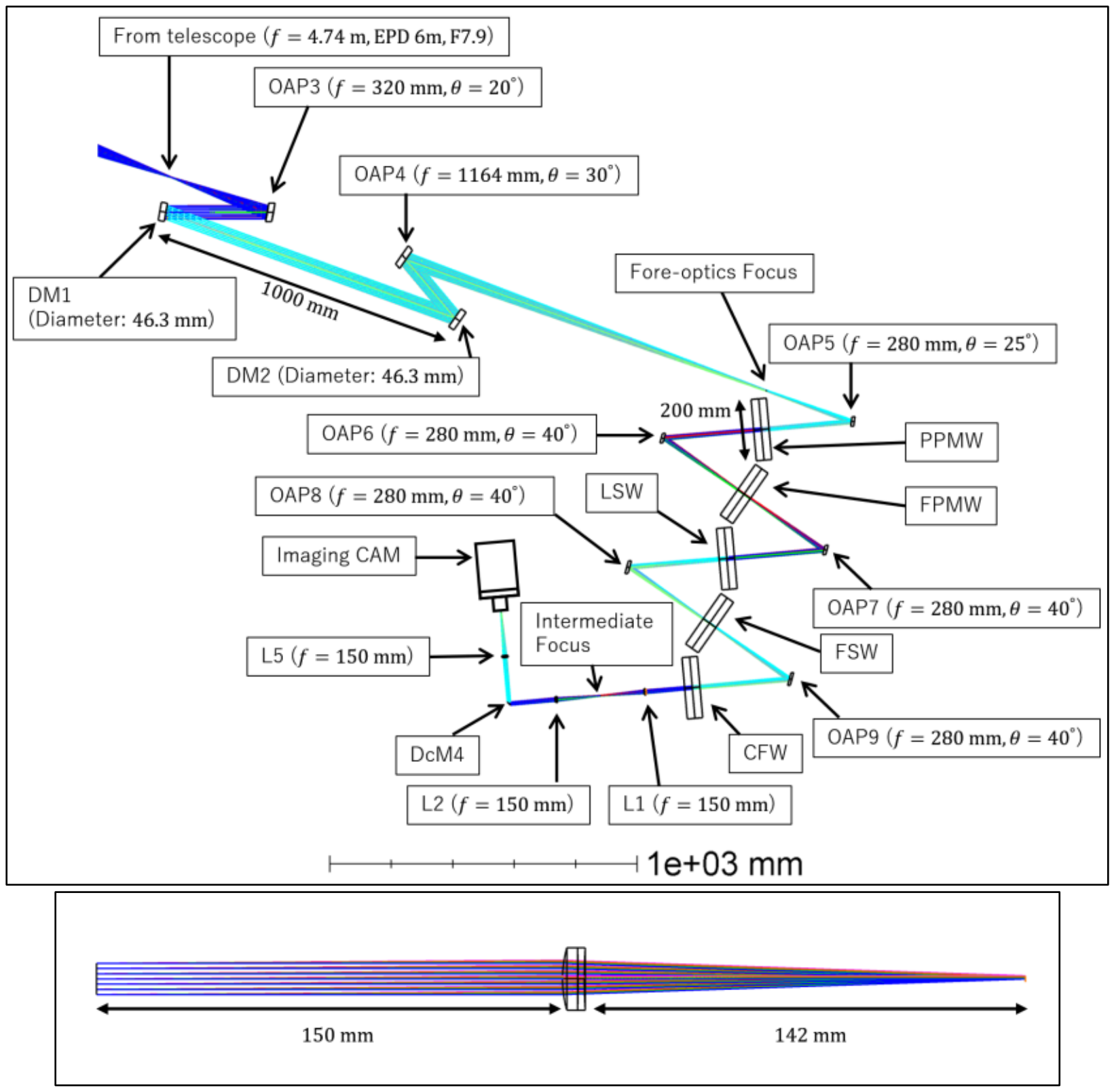

Fig. A1. — Example of the optical design for the NIR coronagraph. (Top) Overall layout of the optical system, where the post-coronagraph section shows only the imaging part and omits the spectroscopic part. (Bottom) Enlarged view of the imaging part.

## Appendix B. Architecture for Coexistence of VIS and NIR Coronagraphs

Fig. A2 presets various architectures for the coexistence of VIS and NIR coronagraphs. This figure is similar to Fig. 5 but provides a more detailed classification of the architectures.

Here, Architecture #4 provides the most comprehensive configuration. It enables simultaneous, ultra-high-contrast (~$10^{-10}$) observation of the same target with wavefront correction in both the VIS and NIR bands. While Architecture #3 is similar to Architecture #4, its configuration is slightly simpler. Architecture #2 is even simpler. Although it allows just to see the same target simultaneously in both the VIS and NIR bands, ultra-high-contrast observation cannot be applied to both simultaneously due to the limitation of effective wavelength range for wavefront correction. In particular, Architecture #4 demands significant resources, and Architectures #2, #3, and #4 require high-performance pre-coronagraph dichroic mirrors. Architecture #1 is the most compact solution shown in Fig. A2. While it does not aim for simultaneous VIS and NIR observations, it is efficient in that it enables simultaneous two-channel (X- and Y-polarization) observations not only in VIS but also in NIR. Architecture #5 is the solution that offers the highest independence between the VIS and NIR channels in Fig. A2.

Simultaneous observation of the same target in VIS and NIR is particularly effective when the required exposure times for both are comparable. Each architecture has its own advantages and disadvantages. Therefore, a systematic trade-off study is effective.

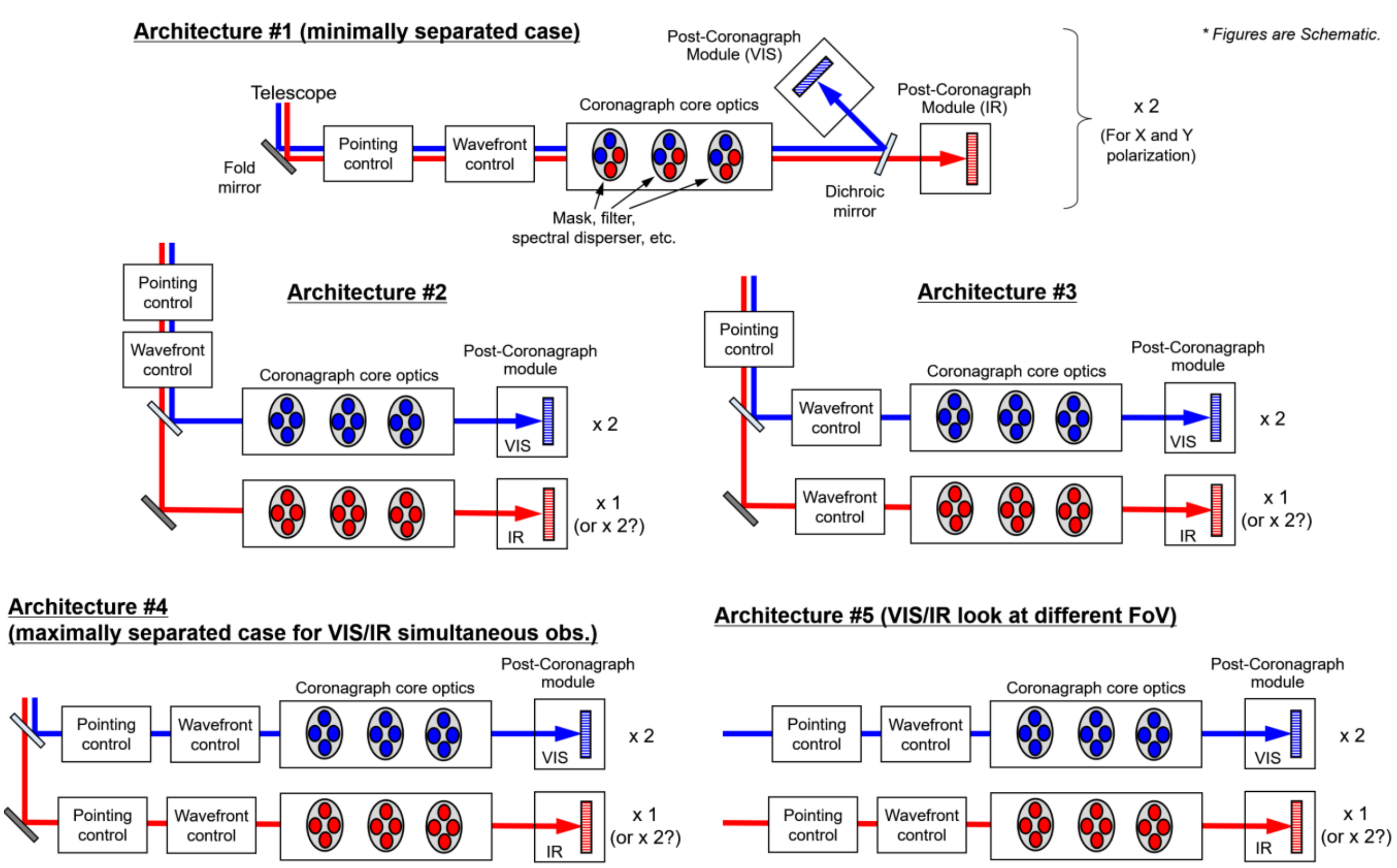

Fig. A2. — Various architectures for the coexistence of VIS and NIR coronagraphs. This figure is similar to Fig. 5 but provides a more detailed classification of the architectures.

## Appendix C. Sensitivity Calculations

We performed sensitivity calculations for the NIR and VIS coronagraph, assuming target star candidates imported from the HWO ExEP Precursor Science Stars Table in the NASA Exoplanet Archive. Our rationale for conducting these calculations, despite the existence of prior studies (e.g., Mennesson et al. 2024), is to allow ourselves the flexibility to adjust parameters and conditions, perform an independent cross-check, and get our technical expertise.

Fig. A3 shows a portion of the calculation results. For targets at very small distances, the calculated exposure time is short, and therefore its wavelength dependence does not constitute a critical problem. On the other hand, these calculations indicate that NIR requires a longer exposure time than VIS for many targets at larger distances. These results can affect the coexisting architecture of VIS and NIR coronagraph hardware and the observation operation plans.

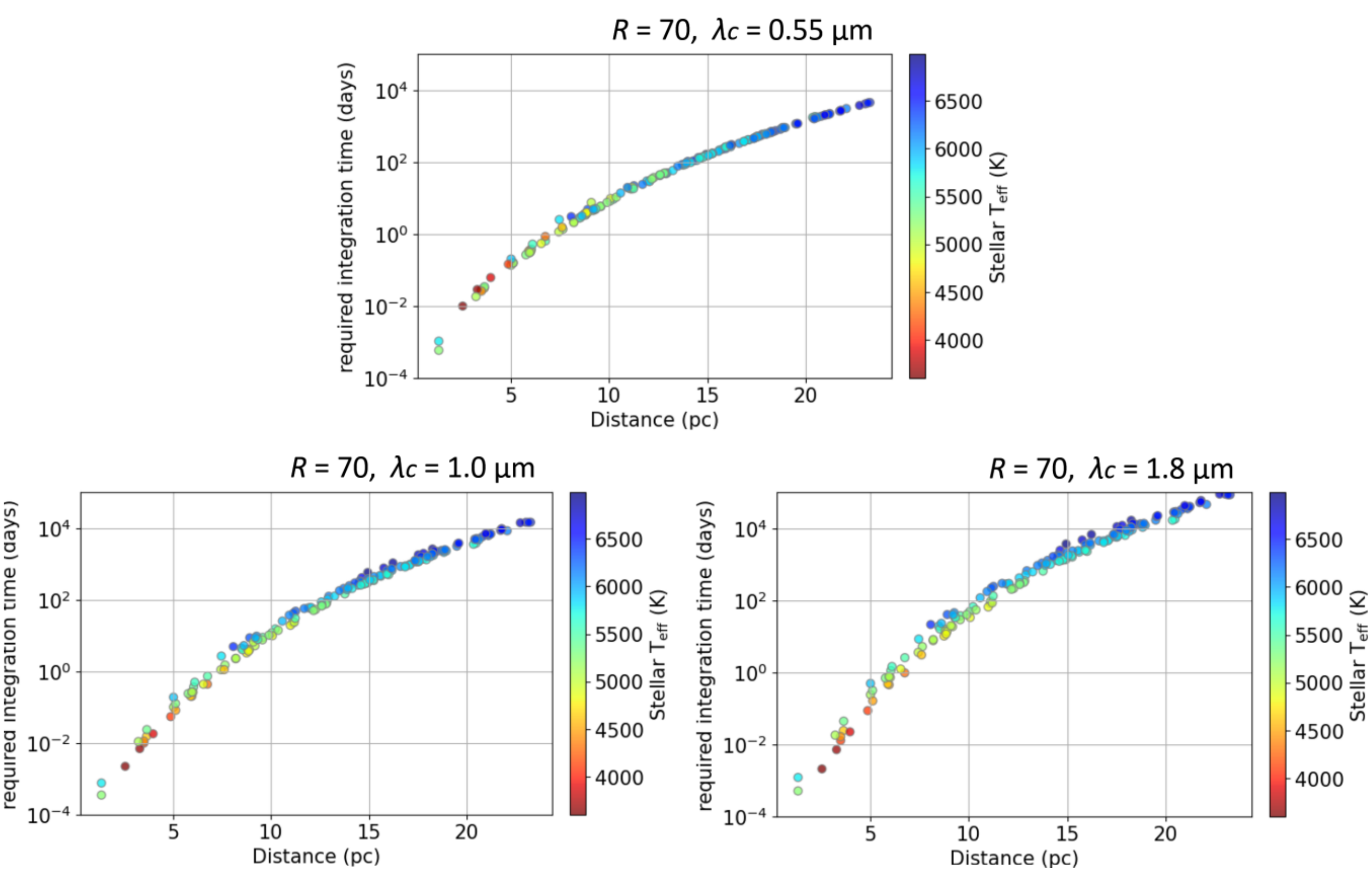

Fig. A3. — Calculated exposure times in VIS and NIR. The vertical axes in the panels indicate the calculated integration time required to reach S/N = 10 for possible planets in the habitable orbits of the assumed target stars, with the assumption shown in the gray rectangle in the upper left. The horizontal axes of the panels show how distant the assumed target star candidates are from the observatory. The colors of the plotted markers mean the target star's effective temperatures.